\documentclass[conference]{IEEEtran}
\IEEEoverridecommandlockouts
\usepackage{cite}
\usepackage{amsmath,amssymb,amsfonts}
\usepackage{algorithmic}
\usepackage{graphicx}
\usepackage{enumitem}
\usepackage{textcomp}
\usepackage{xcolor}
\def\BibTeX{{\rm B\kern-.05em{\sc i\kern-.025em b}\kern-.08em
    T\kern-.1667em\lower.7ex\hbox{E}\kern-.125emX}}

\usepackage{listings}
\usepackage{enumitem}
\usepackage{caption}
\newcounter{listcounter}
\renewcommand{\thelistcounter}{\arabic{listcounter}}

\newcommand{\jecaption}[2]{%
    \vspace{-0.5em} 
    \refstepcounter{listcounter} 
    \begin{center}
    \footnotesize List \thelistcounter: #1
    \label{#2}
    \end{center}
}
\begin{document}

\title{A System to Automatically Generate Configuration Instructions for Network Elements from Network Configuration Models}

\author{
    \IEEEauthorblockN{Nagi Arai}
    \IEEEauthorblockA{
        \textit{Shinshu University}\\
        Nagano, Japan \\
        21w2002h@shinshu-u.ac.jp
    }
    \and
    \IEEEauthorblockN{Shinpei Ogata}
    \IEEEauthorblockA{
        \textit{Shinshu University}\\
        Nagano, Japan \\
        ogata@cs.shinshu-u.ac.jp
    }
    \and
    \IEEEauthorblockN{Hikofumi Suzuki}
    \IEEEauthorblockA{
        \textit{Shinshu University}\\
        Nagano, Japan \\
        h-suzuki@shinshu-u.ac.jp
    }
    \and
    \IEEEauthorblockN{Kozo Okano}
    \IEEEauthorblockA{
        \textit{Shinshu University}\\
        Nagano, Japan \\
        okano@cs.shinshu-u.ac.jp
    }
}


\maketitle

\begin{abstract}
In preparation for constructing or modifying information networks, network engineers develop configuration procedures for network devices according to network configuration specifications. However, as engineers typically create these procedures manually, the generated configuration procedures frequently diverge from the specified requirements. To improve this situation, this paper proposes a method for automatically generating configuration procedures consisting of network device configuration commands based on network configurations and their modification specifications. In this study, we employed the UML (Unified Modeling Language) object-oriented modeling language to develop a notation for network configuration modeling that ensures both strict specification adherence and ease of extension. Additionally, we implemented a method for automatically generating configuration procedures that match the specifications by utilizing network configuration models. As an evaluation experiment, we applied the proposed method to a configuration change scenario in a wide-area campus network at Shinshu University, where the network was migrated from static routing to dynamic routing using the OSPF protocol. As a result, all expected configuration procedures were obtained and a network exhibiting the intended behavior was successfully constructed.
\end{abstract}


\section{Introduction}
In preparation for constructing or modifying information networks, network engineers create configuration procedures for network devices after carefully reviewing their specifications \cite{ipa,miyata}. These procedures—which specify configuration commands for network devices—are designed to ensure consistent configuration practices independent of engineer expertise while enabling later verification of configuration details. Since engineers typically create these procedures manually based on network configuration specifications and their modifications, a common issue arises where generated procedures often deviate from the actual specifications.

Directly contributing to this improvement are previous studies that investigate automated generation of configuration procedures based on network configuration specifications and their modifications \cite{Yamaura20iot,hitachi}. This approach proves effective for obtaining procedures that are consistent with the specified requirements. However, accurately automating the generation of complex and diverse configuration procedures presents challenges in terms of enhancing both the precision and expressiveness of the underlying specifications. Regarding precision improvements, existing research has explored approaches utilizing models—semi-formal specification descriptions incorporating diagrammatic notations \cite{Huang,vsUML,DSML,DSL,MDE,MDN}. Nevertheless, these methods only cover specifications currently available, with no consideration given to representation methods that account for future extensions.

Other relevant approaches include automated network configuration management using tools like Ansible and NETCONF. With regard to Ansible specifically, network and device configuration states are documented using YAML, enabling comprehensive specification management. Furthermore, from these specifications, documentation such as network configuration procedures can be generated using Playbooks. However, Ansible-based network configuration changes assume prior understanding of the grammar structures of network devices, making it difficult for any engineer to easily adapt to various network modifications. Additionally, specifying command-based requirements each time network configurations are managed places significant burden on engineers.

While enhancing specification extensibility is important, merely covering currently available communication protocols would be transient. We believe the key challenge lies in developing specification notations that can be easily maintained and managed by engineers as communication protocols evolve. However, no established methods currently exist for achieving both specification precision and extensibility, nor for automatically generating configuration procedures based on such specifications.

To address this, this study proposes a novel modeling notation for network configuration that applies the Object-Oriented Modeling Language UML (Unified Modeling Language). This notation-based model can provide detailed information at the parameter value level of network device configuration commands. UML is a graphical notation system consisting of a single metamodel, primarily used for describing and designing systems built using object-oriented methodologies \cite{uml1, uml2}.

Typically, when constructing or modifying information networks, designers comprising various engineers create network configurations, with the primary role of network engineers being to develop detailed designs for network construction and configuration changes. These designs must be instructions that network engineers can follow without hesitation and with maximum clarity. Focusing on UML, we propose a modeling notation that visually represents complex network configuration concepts and relationships, while providing more specific and granular details through class and object descriptions. This approach aims to enhance both the extensibility and precision of network specifications. Furthermore, this study also newly implements a system that automatically generates device configuration procedures for each network device based on differences between two network configuration models. This method enables precise automatic generation of complete sequences of network device configuration commands from detailed configuration model differences, while also implementing a new notation for describing correspondence relationships between models and configuration commands (hereafter referred to as "device configuration command templates"). This facilitates easier extension and modification of generated configuration commands by engineers. By enabling both network system design and automatic generation of appropriate configuration commands to be issued to network devices, this approach is expected to significantly reduce engineers' workload.

As part of our evaluation experiments, we applied the proposed method to a configuration change scenario involving multiple network devices in an actual operational network using the OSPF protocol. The results confirmed that all configuration procedures were obtained as expected, and a network exhibiting the intended behavior was successfully constructed.

The paper is organized as follows: Section 2 describes the proposed method, with Section 3 explaining the generated device configuration procedures. Section 4 discusses the evaluation, summarizing the assessment conducted to verify the validity of the proposed method and presenting the results. Finally, Section 5 provides conclusions and outlines future research directions.
\begin{figure*}[!t]
        \centering
        \includegraphics[width=0.9\hsize]{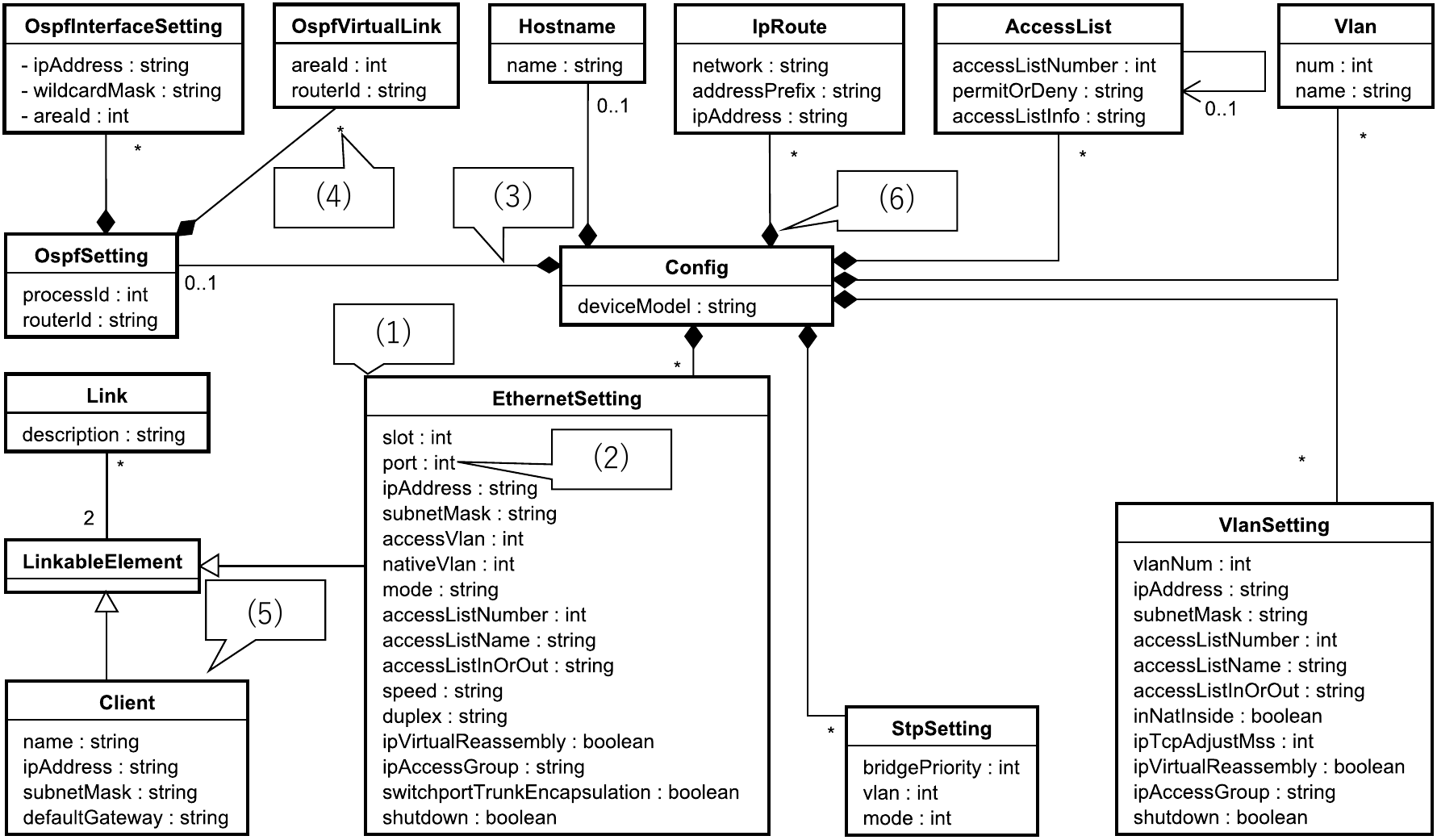}
        \caption{Network configuration metamodel}
        \label{fig_nw_metamodel}
\end{figure*}

\begin{figure}[!t]
        \centering
        \includegraphics[width=0.9\hsize]{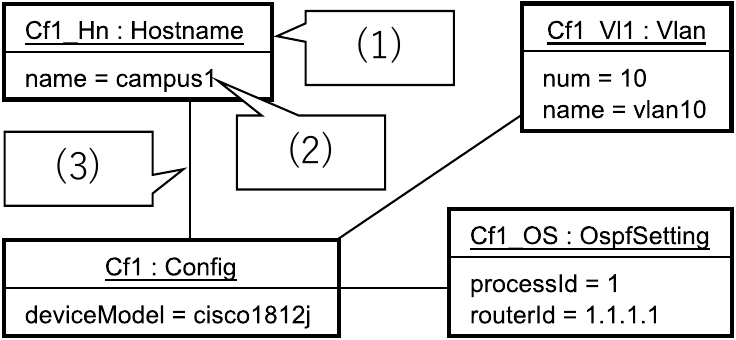}
        \caption{Network configuration model}
        \label{fig_nwm_example}
\end{figure}
\section{Proposed Method}
The proposed method is explained in two parts: the first covers the network configuration metamodel that defines the grammatical structure of network configurations, and the network configuration model that represents instances of network configurations. The second part describes the method for automatically generating device configuration procedures based on differences between two network configuration models and device configuration command templates.

\subsection{Network Configuration Metamodel and Model}
Network configuration information encompasses both the physical configuration of network devices (such as cabling) and the settings applied to each device. In operational networks, clearly documenting network configuration information and thoroughly reviewing it among engineers becomes crucial to prevent construction and configuration change failures. To clearly define network configuration information during the design phase, it is essential to represent the essential data—including parameter values for configuration commands—at their native granularity, while also employing notational conventions that facilitate organization of these data.

Conventional network configuration diagrams \cite{miyata} typically depict network configurations by connecting icons representing network devices and other elements with lines. While this approach makes the configuration structure clear, it often leaves insufficient space for detailed device configuration specifications and tends to become ambiguous. Furthermore, even when detailed device configuration information is optionally included, it must be presented in free-form text, making it highly dependent on the author's skill level and difficult to guarantee quality.

As a method aimed at enhancing information content and formalizing documentation, Yoshizawa et al.'s system \cite{hitachi} enables both visualization of network configuration diagrams and display of dialog boxes for entering detailed network device configurations. Consequently, it allows for more rigorous and detailed management of network configurations compared to conventional diagramming methods. However, expanding or modifying network device configuration items requires software modifications, which generally involve significant costs. If the system's data structure specifications could be formally documented, this would facilitate easier configuration item expansion, though no approach for generating configuration procedures based on specification documentation has been presented.

Therefore, in this research, we propose applying the UML object-oriented modeling language to develop a notational method for network configuration modeling that balances both rigor and extensibility. The primary conceptual elements in our proposed notation include the network configuration metamodel (hereafter simply referred to as "metamodel") for structurally defining configuration specification items, and the network configuration model (hereafter simply referred to as "model") for representing specific network configurations. For example, while the metamodel specifies configuration items like "port," the model would define details such as "port" with specific values like "2". The following sections provide detailed explanations of both the metamodel and model.

\begin{figure*}[!t]
        \centering
        \includegraphics[width=0.9\hsize]{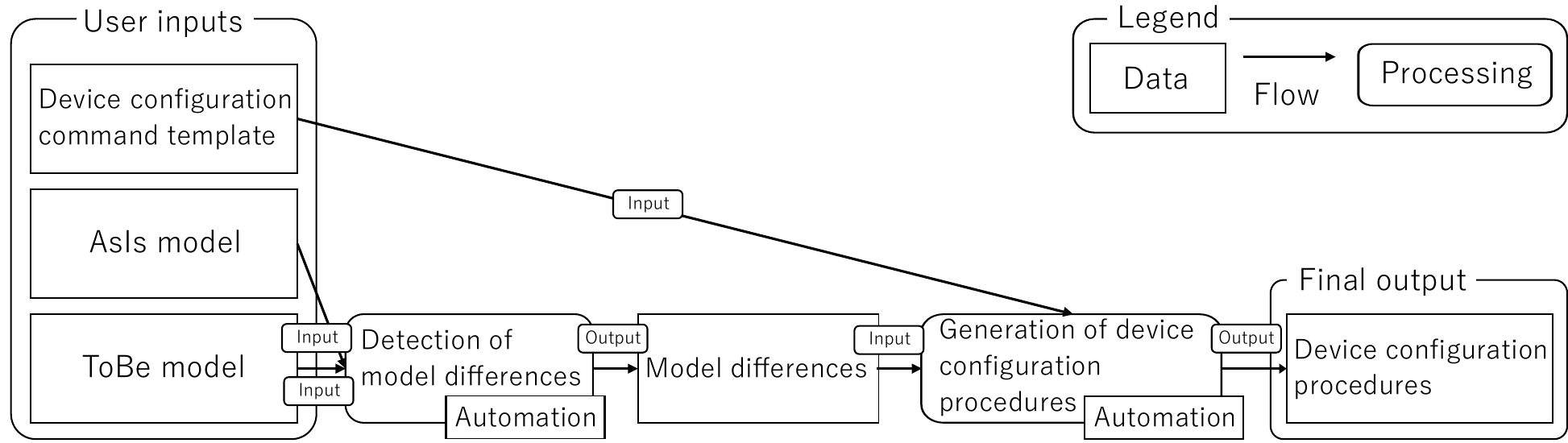}
        \caption{Overview of the proposed method to generate configuration commands for network elements}
        \label{fig_method}
\end{figure*}

\begin{figure*}[!t]
        \centering
        \includegraphics[width=0.95\hsize]{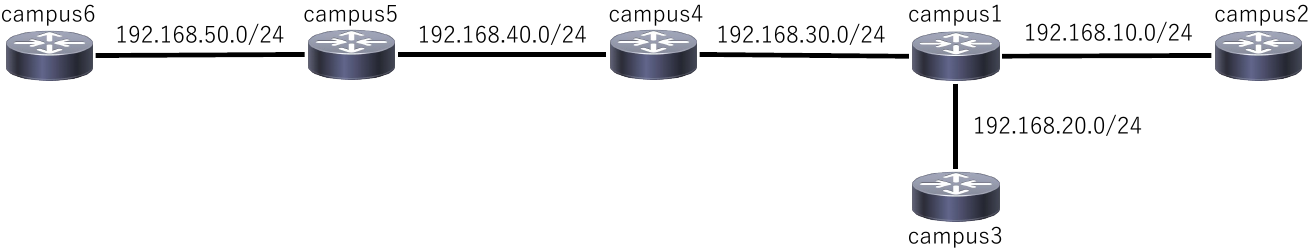}
        \caption{AsIs network diagram}
        \label{fig_asis_nw}
\end{figure*}

\begin{figure*}[!t]
        \centering
        \includegraphics[width=0.9\hsize]{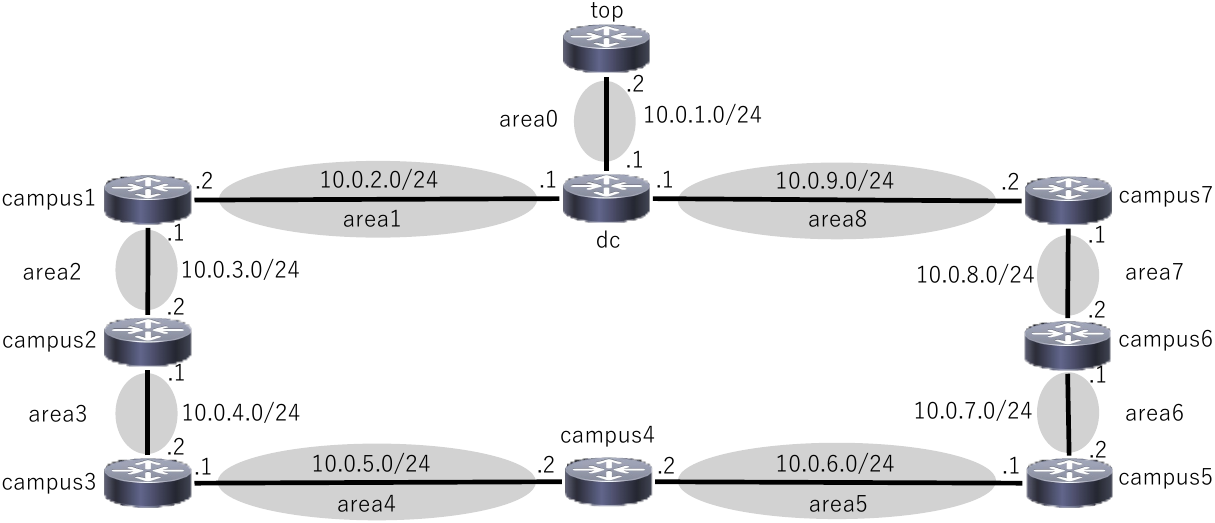}
        \caption{ToBe network diagram}
        \label{fig_tobe_nw}
\end{figure*}

\subsubsection{Network Configuration Metamodel}
Fig. \ref{fig_nw_metamodel}. shows the metamodel relevant to this document, presented using UML class diagram notation. The main constituent concepts are numbered 1 through 6 in the figure and are summarized below:

\begin{enumerate}
    \item Groups of closely related specification items, such as "ipAddress" and "port" (e.g., the large rectangular area labeled "EthernetSetting"), where "EthernetSetting" serves as the group name.
    
    \item Individual specification items representing items to be specified (e.g., "port:int"). In this example, "port" is the item name and "int" indicates the data type.

    \item Relationships between specification item groups (e.g., the line connecting "OspfSetting" and "Config").

    \item At the endpoints of these relationships, multiplicity notations appear (e.g., "*"), where "*" means "zero or more". For instance, when there is one OspfSetting, it implies there can be zero or more related OspfVirtualLinks. Note that "0..1" means "zero or one".

    \item Relationships marked with white triangles are called generalizations, meaning that the properties on the triangle side are inherited by the unmarked side. This semantics allows, for example, considering a "Client" to have a relationship with "Link".

    \item Relationships marked with black diamonds are called composition, meaning that the diamond-side entity owns the unmarked-side entity. This semantics prevents, for example, a situation where a single OspfSetting could share multiple OspfInterfaceSettings.

\end{enumerate}

Thus, the metamodel in this study defines a combination of specification item groups from the physical design perspective (e.g., "Link" and "LinkableElement") and those from the logical design perspective (e.g., "EthernetSetting" and "OspfSetting"). This approach enables consistent organization of specification item structures from different perspectives using a single notation.

Moreover, most of the logical design specification items in this paper are defined based on the command syntax of Cisco network devices. While the number of specification items shown in Fig. \ref{fig_nw_metamodel} is not extensive, extending this framework consistently with the command system is easy. On the other hand, while this paper's metamodel does not aim for multi-vendor compatibility, compatibility with devices from other vendors that maintain high consistency with Cisco's command syntax would be relatively straightforward to implement.

\subsubsection{Network Configuration Model}
Fig. \ref{fig_nwm_example}. illustrates the model, which is presented using the UML object diagram notation. The model adheres to the metamodel shown in Fig. \ref{fig_nw_metamodel}. This compliance refers to the following: (1) through (3) below, which correspond to the notations in the figure.

\begin{enumerate}

\item Creating values (hereafter referred to as "group values"; e.g., the large rectangle labeled Cf1:Config) that are instances of specification item groups from the metamodel. Here, f1:Config is called the identifier, while Cf1 is the group value name. Config is the name of the specification item group.

\item Assigning values (hereafter referred to as "item values"; e.g., campus1 in name=campus1) to each specification item within the group values.

\item Depicting lines between group values (hereafter referred to as "relationship values") that satisfy the relationships between specification item groups.

\end{enumerate}

Note that specification items with empty values do not have any content following the = in entries like name=campus1. Here, the equal sign (=), based on UML object diagram semantics, indicates a state where the right-hand side item value has been assigned to the left-hand side specification item.

\subsection{Device Configuration Procedure Generation Method}\label{sec_gen}
Fig. \ref{fig_method}. outlines the general approach for generating device configuration procedures. This method generates configuration procedures based on differences between two models. These two models are assumed to be in the relationship where one represents the current network configuration that needs modification (AsIs model), and the other represents the desired network configuration after modification (ToBe model). The following describes the method's workflow:

\begin{figure*}[!t]
        \centering
        \includegraphics[width=\hsize]{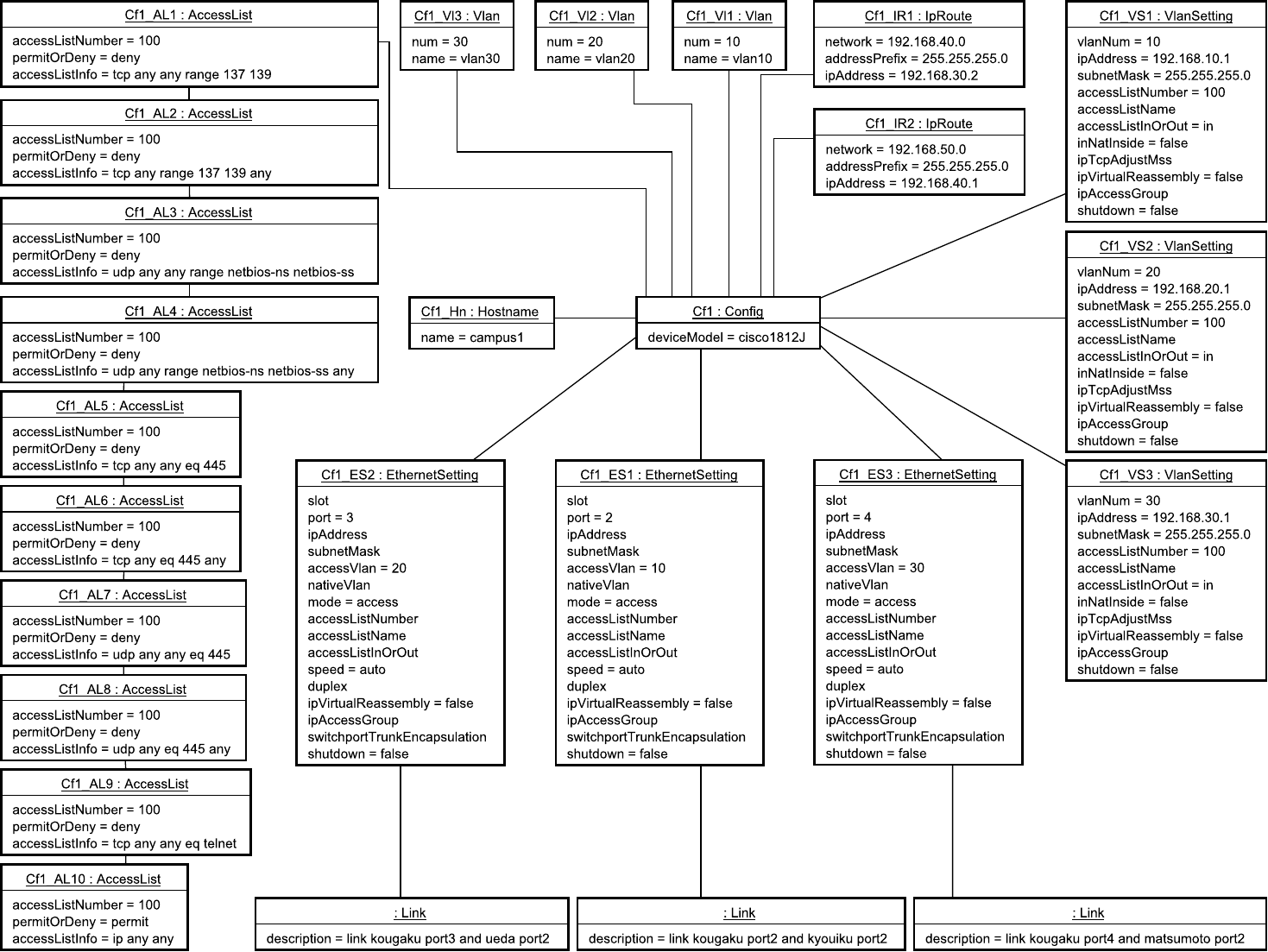}
        \caption{AsIs network configuration model (campus1 in Fig. \ref{fig_asis_nw})}
        \label{fig_asis_model}
\end{figure*}

\begin{figure*}[!t]
        \centering
        \includegraphics[width=0.95\hsize]{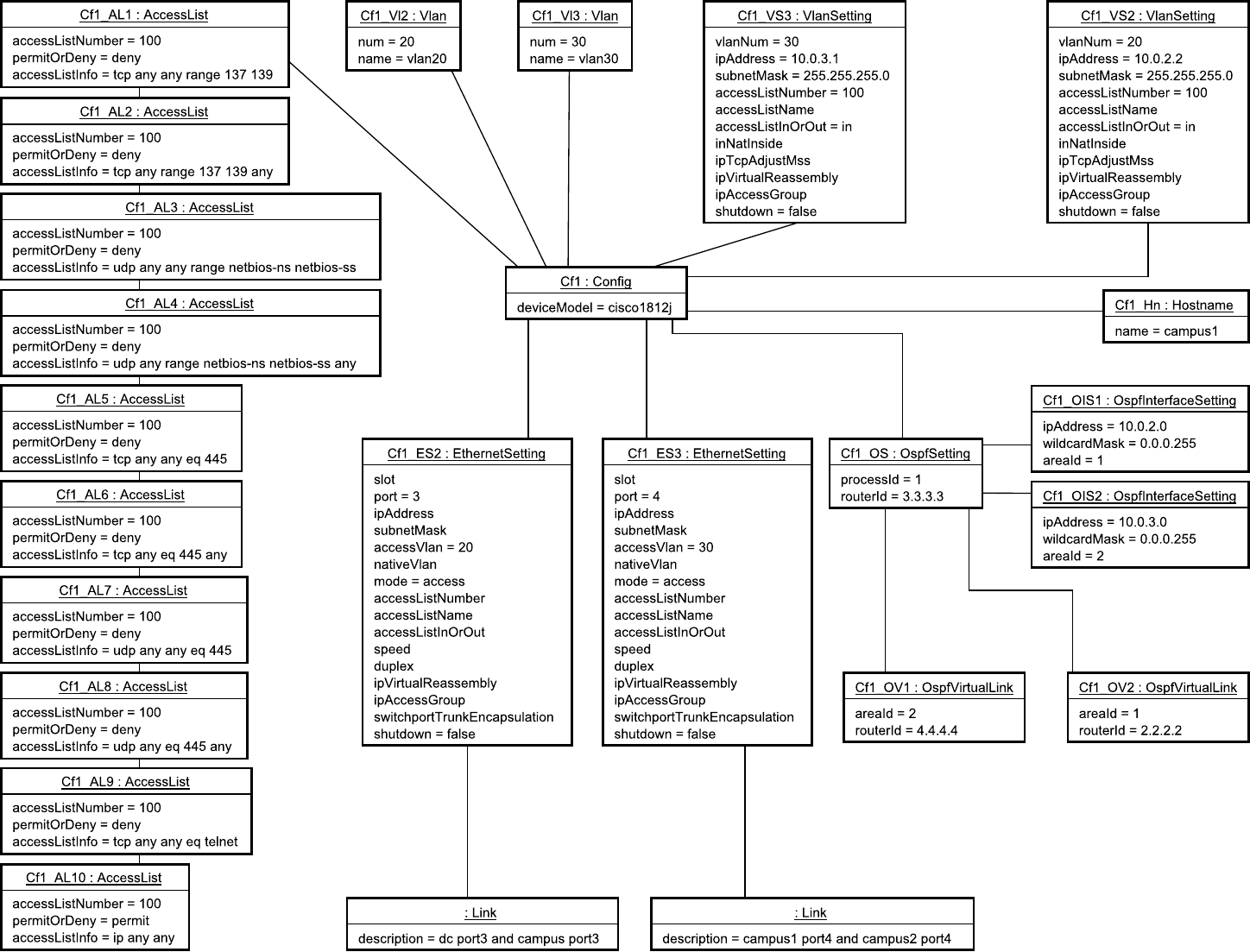}
        \caption{ToBe network configuration model (campus1 in Fig. \ref{fig_tobe_nw})}
        \label{fig_tobe_model}
\end{figure*}

\begin{figure*}[!t]
        \centering
        \includegraphics[width=0.95\hsize]{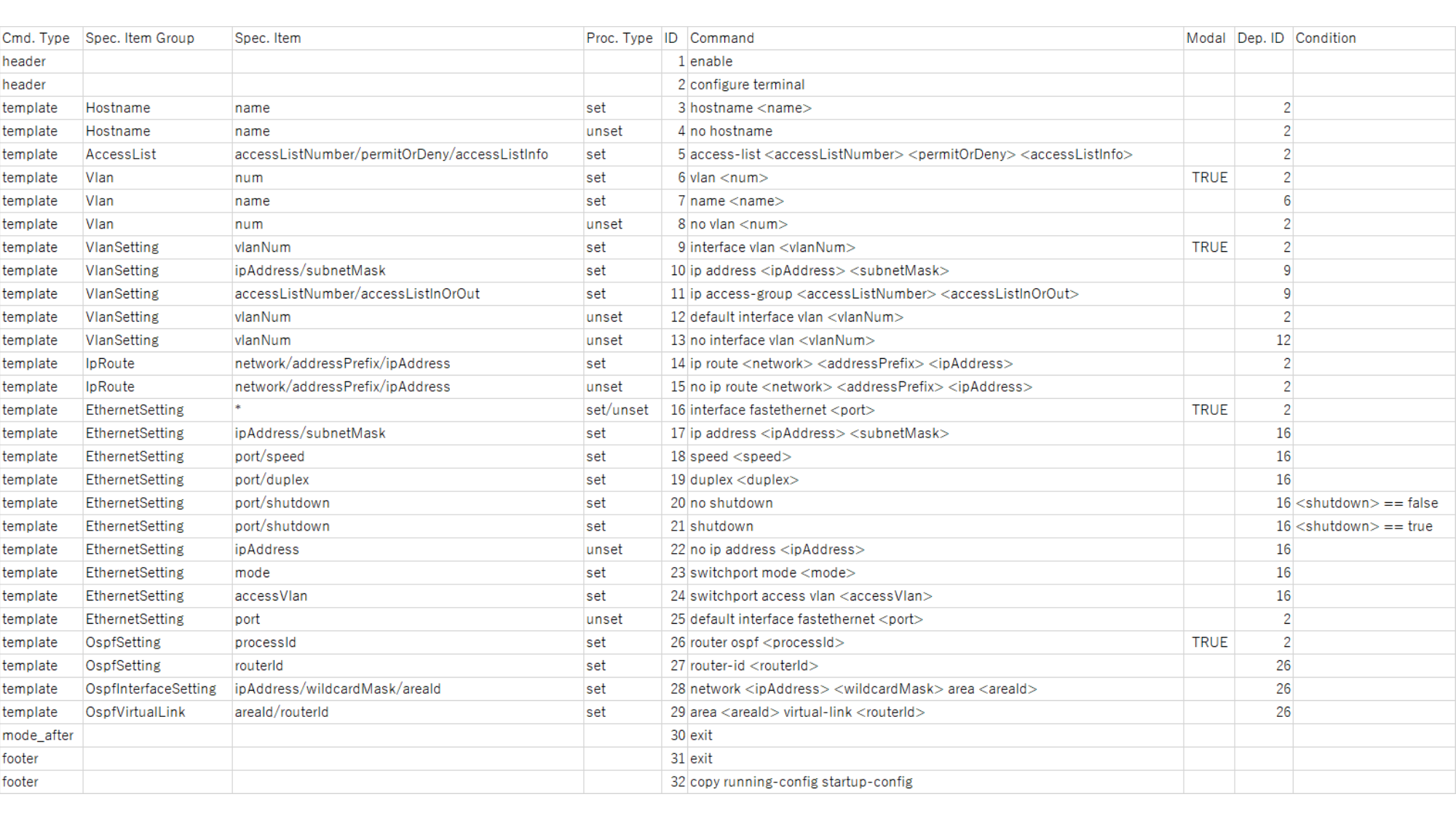}
        \caption{Template of configuration commands for network elements (based on Cisco 1812-J and Cisco 892)}
        \label{fig_template}
\end{figure*}

\begin{description}
\item [1.User Inputs] This method requires three inputs from the user: the AsIs model (Fig. \ref{fig_asis_model}), the ToBe model (Fig. \ref{fig_tobe_model}), and the device configuration command template (Fig. \ref{fig_template}), with the final output being the device configuration procedures. These configuration procedures refer to the set of device configuration commands obtained for each network device required to transform the AsIs configuration into the ToBe configuration. One key reason for generating procedures based on model differences is that this enables easier rollback in case of any issues in the applied network. While it would also be possible to generate configuration procedures directly from the AsIs and ToBe models, this approach would lack the ability to reference the differences from the original network configuration, making it more difficult to perform rollbacks. Additionally, when considering changes to interfaces with ACLs (Access Control Lists) or remote operations, it is preferable to generate configuration procedures only for the specific areas requiring modification.

\item [2.Detection of Model Differences] This step involves identifying the differences between AsIs and ToBe models. Model differences essentially represent combined information about configuration settings that need to be removed and those that need to be newly applied. By obtaining these model differences, we can determine the specific device configuration commands required to transform the AsIs configuration into the ToBe configuration. The procedure for detecting model differences is detailed in Section \ref{sec_sabun}.

\item [3.Generation of Device Configuration Procedures:]This step generates the device configuration procedures by concretely implementing the device configuration command template based on the model differences. The procedure for generating configuration procedures is detailed in Section \ref{sec_generate}.

\end{description}

As a demonstration example of this method, we will consider the case where changing the network configuration from Fig. \ref{fig_asis_nw} to Fig. \ref{fig_tobe_nw}. For space reasons, only partial representations of the AsIs and ToBe models are shown. This example is based on a simulated network configuration from Shinshu University \cite{Suzuki_1}, involving the configuration change from a static routing network to a dynamic routing network using the OSPF protocol. The corresponding models for Fig. \ref{fig_asis_nw} and Fig. \ref{fig_tobe_nw} are shown in Fig. \ref{fig_asis_model} and Fig. \ref{fig_tobe_model}, respectively.

\subsubsection{Detection of Model Differences}\label{sec_sabun}
Model differences refer to discrepancies in item values between the AsIs and ToBe models. When there are differences in item values, the values in the AsIs model require removal from the network device, while the values in the ToBe model require new configuration to be applied to the network device. The objective of this step is to assign an "unset" label to items that need to be removed and a "set" label to items that need to be newly configured.

Next, we describe the algorithm for detecting model differences. As an initial assumption, item values for empty fields (i.e., unused items) must always be unlabeled. Therefore, labeling operations described below are performed only on non-empty item values. First, we pair group values that share the same identifiers (e.g., Cf1\_Hn or Cf1\_Vl1) between the AsIs and ToBe models. When a group value pair is established, if there are discrepancies in item values between items with the same name within the pair, the AsIs side item value receives an "unset" label while the ToBe side item value receives a "set" label. If there are no items with the same name due to differences in model structures between AsIs and ToBe, the item value present only in the AsIs side receives an "unset" label, while the item value present only in the ToBe side receives a "set" label. If no group value pair is established, an "unset" label is assigned to all item values belonging to the group value on the AsIs side, and a "set" label is assigned to all item values on the ToBe side. By assigning labels in this way—whether for new configuration/update or removal—we can obtain the necessary information for generating the appropriate commands. Note there are no specific guidelines for the exact format of identifiers.

As shown in Fig. \ref{fig_asis_model} and Fig. \ref{fig_tobe_model}, label assignments have been made for item values that should not be configured in network devices (e.g., item values for the "description" field within the "Link" structure representing wiring information). The "Link" and "Client" entities mentioned here represent groups of specification items that describe connections to network devices or clients, and no configuration commands are involved in generating device configuration procedures. Therefore, even if labels are assigned before or after changes, they should be disregarded. Such item values should be excluded from the final output by the device configuration procedure generation algorithm or configuration command template.

\begin{table*}[tb]
\begin{center}
    \caption{Components of the device configuration command template}
    \small
    \begin{tabular}{l|p{14cm}} \hline \hline
    \label{tab_template_terms}
    Item name & Description \\ \hline
    {\tt Cmd. Type} & Represents the type of device configuration command.
        Specifically, the following types exist:
        \begin{description}
           \item[{\tt template}] Represents a device configuration command whose necessity is determined according to the labels of item values.
           \item[{\tt header}] Represents device configuration commands that must be executed at the beginning of device configuration.
           \item[{\tt footer}] Represents device configuration commands that must be executed at the end of device configuration.
           \item[{\tt mode-before}] Represents device configuration commands that must be executed immediately before a mode is changed. Here, a mode refers to a state that determines which configuration items can be configured, and a mode change refers to switching from one such state to another.
           \item[{\tt mode-after}] Represents device configuration commands that must be executed immediately after all device configuration commands required in the current mode have been issued.
        \end{description}
        Not all of the above types must necessarily be used; for example, in Fig.\,\ref{fig_template}, no {\tt mode-before} commands are specified.\\ \hline
    {\tt Spec. Item Group} & Represents a specification item group.
        In this context, a specification item group is one of the pieces of information used to decide which device configuration commands are required.
        For example, if this element has the value {\tt Hostname}, it means that the device configuration command in the same row is required when the item value of a particular specification item in {\tt Hostname} has a specified label.
        This element is only applicable to {\tt template} commands. \\ \hline
    {\tt Spec. Item} & Represents a specification item.
        In this context, a specification item is one of the pieces of information used to decide which device configuration commands are required.
        For example, if this element has the value {\tt port/shutdown}, it means that the device configuration command in the same row is required when the item value of either {\tt port} or {\tt shutdown} has a specified label.
        The symbol {\tt *} is an abbreviation meaning ``any specification item belonging to the {\tt Spec. Item Group}''.
        This element is only applicable to {\tt template} commands. \\ \hline
    {\tt Proc. Type} & Represents a label indicating whether a new configuration is required {\tt set}, a configuration must be removed {\tt unset}, or either of the two {\tt set/unset}.
        For example, if this element has the value {\tt set}, it means that the device configuration command in the same row is required when the label of the item value is {\tt set}.
        This element is only applicable to {\tt template} commands. \\ \hline
    {\tt ID} & Represents a number that identifies the device configuration command. \\ \hline
    {\tt Command} & Represents a device configuration command. Here, a device configuration command can abstract certain parts using specification item names from the metamodel (e.g., {\tt hostname <name>}).
        A string enclosed in {\tt <} and {\tt >} (e.g., {\tt <name>}) denotes a specification item name.
        When a device configuration command is concretized, {\tt <}specification item name{\tt >} (e.g., {\tt <name>}) is replaced with the corresponding item value (e.g., {\tt Router1}).
        This element is only applicable to {\tt template} commands. \\ \hline
    {\tt Modal} & Indicates whether the device configuration command causes a mode change.
        If a mode change occurs, this field is written as {\tt TRUE}; if no mode change occurs, it is left blank.
        This element is only applicable to {\tt template} commands.\\ \hline
    {\tt Dep. ID} & Represents, by {\tt ID}, the device configuration command (priority command) that must be executed beforehand.
        This imposes constraints on the order in which commands are issued.
        For example, a device configuration command that requires a prior mode change is given the {\tt ID} of the device configuration command that performs that mode change.
        If no priority command exists, this field is left blank.\\ \hline
    {\tt Condition} & Represents a condition on item values used to decide whether a device configuration command is required. If this field is empty, it is treated as always true.
        When the condition is not satisfied, the corresponding device configuration command is regarded as unnecessary.
        The condition notation uses only the equality operator ({\tt ==}); for example, {\tt <}specification item name{\tt > == }item value (e.g., {\tt <shutdown> == true}).
        This element is only applicable to {\tt template} commands. \\ \hline \hline
    \end{tabular}    
    \\ Note: The full names of abbreviations are as follows. Cmd. Type = Command Type; Spec. Item Group = Specification Item Group;\\ Spec. Item = Specification Item; Proc. Type = Process Type; Dep. ID = Dependent ID.
\end{center}
\end{table*}

\begin{figure*}[!t]
        \centering
        \includegraphics[width=\hsize]{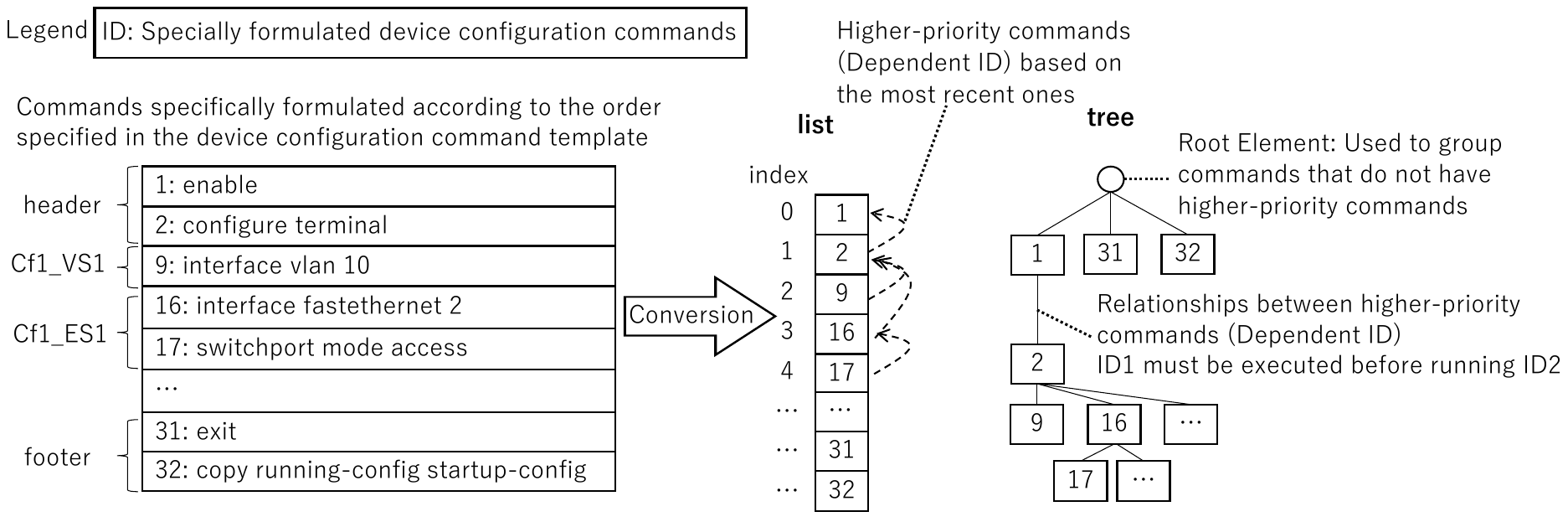}
        \caption{Processing of header, template, and footer commands}
        \label{fig_com_process}
\end{figure*}

\begin{figure}[!t]
        \centering
        \includegraphics[width=\hsize]{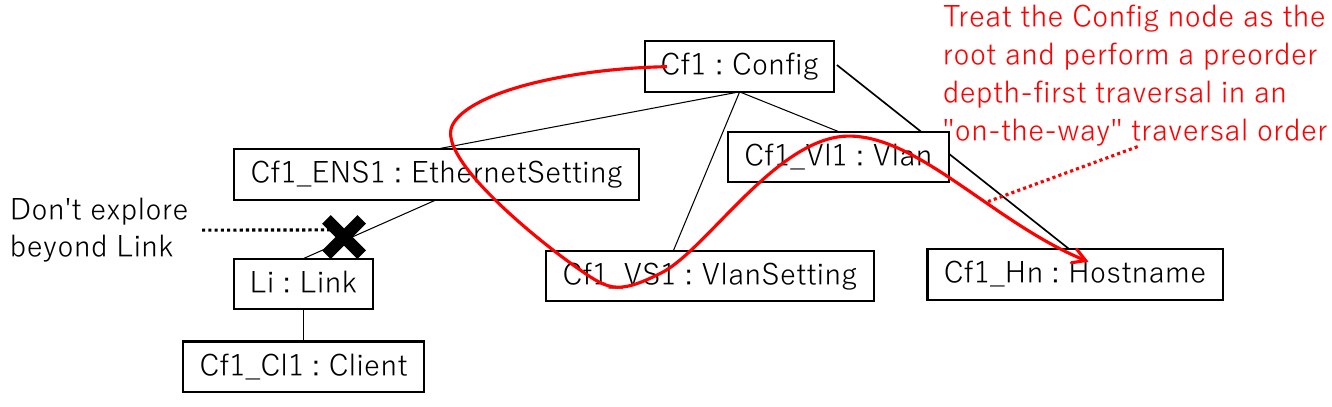}
        \caption{Overview of traversal algorithm for specification item
groups}
        \label{fig_sig_search}
\end{figure}

\subsubsection{Device Configuration Procedure Generation} \label{sec_generate}
To apply the model differences obtained in the previous step to network device configuration, corresponding device configuration commands must be generated based on the labels assigned to each item value.

Fig. \ref{fig_template}. illustrates an example of a device configuration command template, with Table \ref{tab_template_terms} explaining its constituent elements. Since configuration command structures may vary depending on the network device model, we create device configuration command templates specific to each model by matching the "deviceModel" value in Config with the name of the template (not visible in Fig. \ref{fig_template}). Additionally, specification item groups not listed in Fig. \ref{fig_template}'s "Spec. Item Group" (such as "Link" and "Client") should not be used for generating device configuration procedures.

For items with strict ordering constraints such as ACLs (Access Lists), no ordering constraints are applied using configuration command templates. Instead, constraints are imposed on the relationships that appear in the model. For example, in the AccessList shown in Fig. \ref{fig_nw_metamodel}, a self-referencing relationship with a multiplicity of 0 to 1 ensures that each AccessList instance references another instance of the same type, effectively representing a single-directional list. Consequently, the device configuration procedure for an AccessList appearing in the network configuration model is generated by an algorithm that produces the procedure in the order specified by these relationships. Furthermore, when operating a network, modifications such as additions or deletions to ACLs may occur. Generally, two methods are commonly considered for modifying ACLs:

\begin{description}
\item[1.Editing an existing ACL] Each conditional statement in the ACL is assigned a sequence number. This sequence number is a sequentially assigned identifier to each condition defined in the ACL, representing the order in which rules are executed. Typically, sequence numbers are assigned at equal intervals (e.g., Condition A at sequence 10, Condition B at sequence 20, etc.), and these numbers can be used to make modifications to existing ACLs. When adding an ACL, new entries should be inserted between existing entries with evenly spaced sequence numbers. For instance, to add a new condition C between conditions A (sequence 10) and B (sequence 20), you would assign sequence numbers 11 through 19 to condition C, thereby inserting the new entry between the existing conditions. For deletions, you can specify the sequence number of the condition to be removed to delete it. Since sequence numbers can be adjusted at equal intervals, this adjustment should be performed each time a configuration change is made.

\item[2.Creating a new ACL] Another method is to create and apply a new ACL separately from existing ones. Unlike the method described in 1, this approach does not require consideration of sequence numbers. Instead, you create a new ACL that includes all existing conditions along with any new conditions you wish to add.

\end{description}

This methodology adopts the "2. Creating a new ACL" approach. When making configuration changes, we create a new ACL without modifying the existing one, then apply the new configuration instead.

\begin{figure*}[!t]
        \centering
        \includegraphics[width=\hsize]{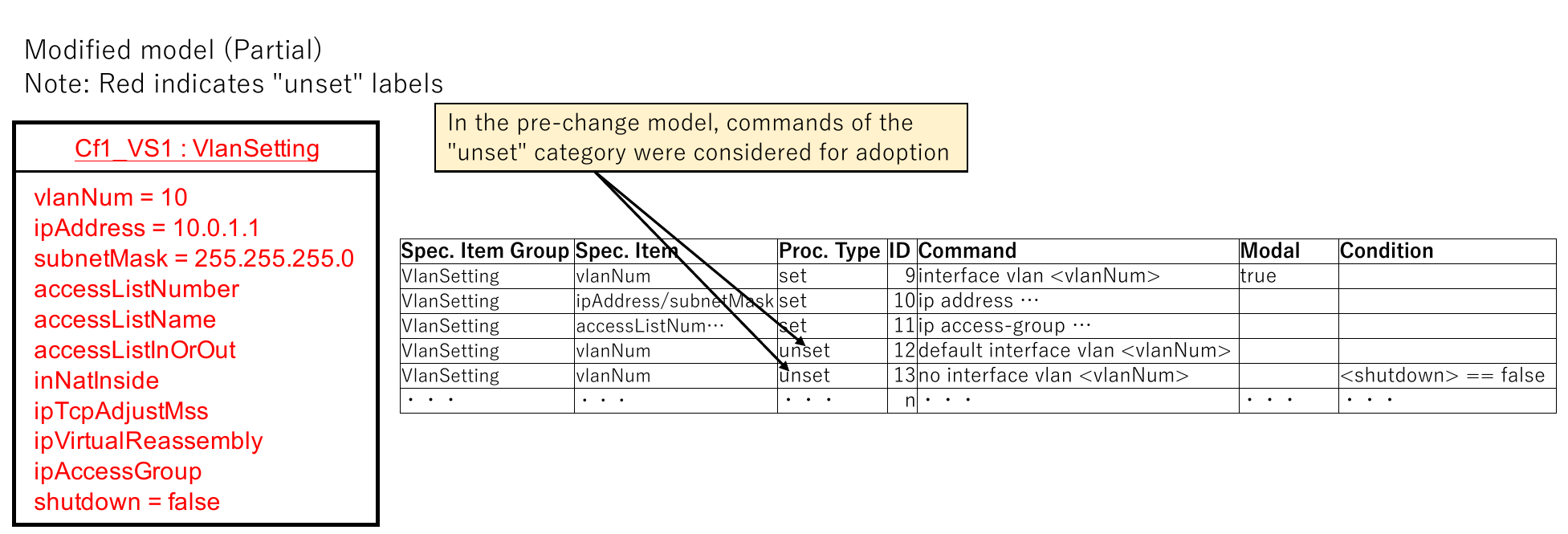}
        \caption{Selection of commands according to changes in specification item groups}
        \label{fig_com_temp_process}
\end{figure*}

\begin{figure*}[t]
        \centering
        \includegraphics[width=\hsize]{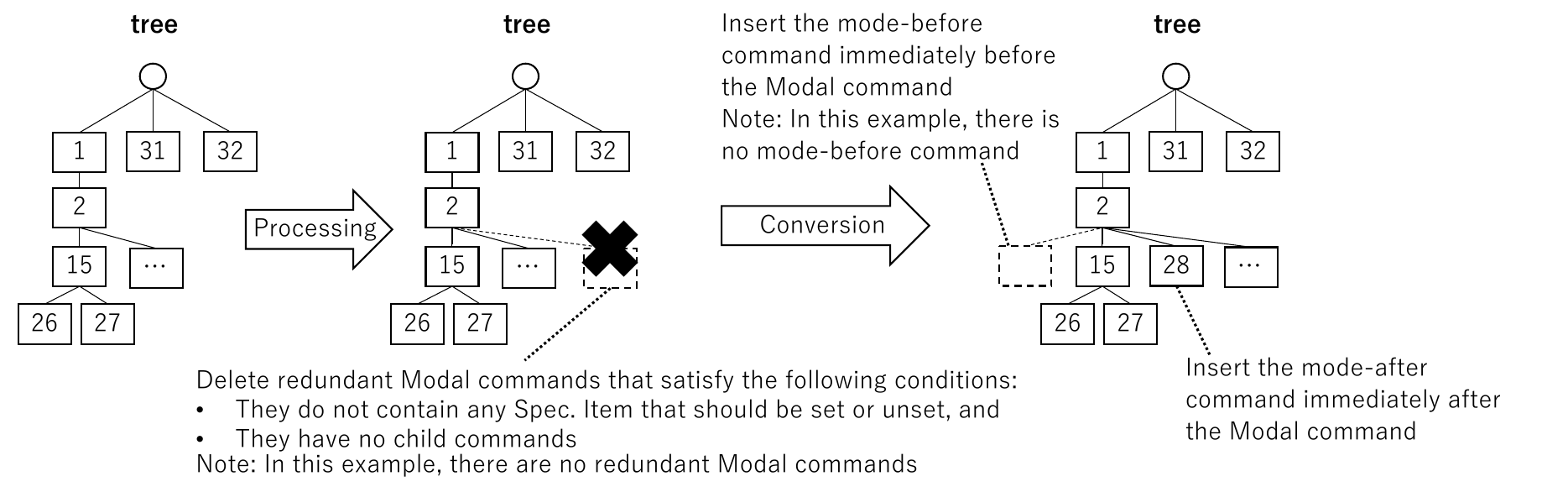}
        \caption{Processing of modal commands and insertion of mode-before and mode-after commands}
        \label{fig_mode_process}
\end{figure*}

\begin{figure}[tb]
        \centering
        \includegraphics[width=0.7\hsize]{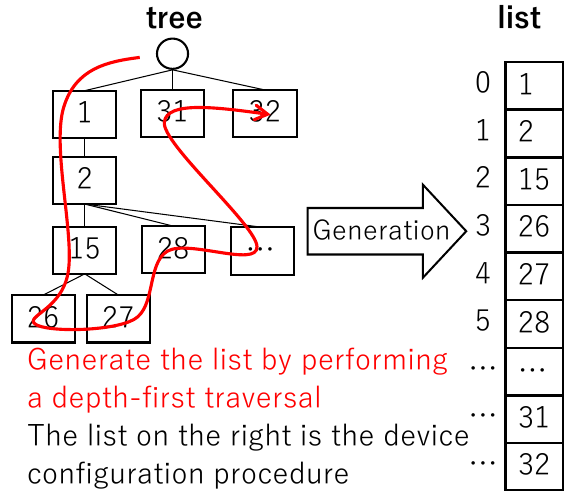}
        \caption{Generation of configuration commands for network elements}
        \label{fig_proc_gen}
\end{figure}

Next, we describe the algorithm for generating device configuration procedures. When a model contains multiple Config group values, the following steps (1) through (7) should be performed for each Config group value separately:

\begin{enumerate}
\item Prepare two data structures: a list that maintains commands in their processing order, and a tree that hierarchically organizes priority command relationships based on Dep. IDs. (Fig. \ref{fig_com_process} on the right illustrates the concepts of list and tree.)

\item As shown in Fig. \ref{fig_com_process}, insert header commands into the list. If any command Ca in the list element has a higher-priority command Cp, reflect the command structure with Cp as the parent and Ca as the child in the tree. In some cases, there may be multiple higher-priority commands Cp1, Cp2 (where ind(Cp1) < ind(Cp2)) within the command Cb in the list element. In this case, for the command Cp2 with the larger index (the most recent one), reflect the command structure with Cp2 as the parent and Cb as the child in the tree.

\item To generate configuration removal commands, process the Config in the AsIs model as follows:

\begin{enumerate}
    \item Treat the model as a tree structure with the Config node as the root and perform a preorder depth-first traversal in an "on-the-way" traversal order. (Fig. \ref{fig_sig_search}) However, when encountering a Link command that represents wiring information to another device, skip processing the Link and return to the search without proceeding further.
    \item For each node reached during traversal (specification item group values), process Command commands of the "unset" category corresponding to the group values in the device configuration command template in top-down order. (Fig. \ref{fig_com_temp_process}) When either of the following conditions is met, concretely specify the Command using the item values and add it to the list; also, if a higher-priority command exists, reflect it in the tree.
    \begin{description}
        \item[Condition 1.] 
        \quad
        \begin{itemize}
            \item If any of the item values associated with the Spec. Item contain an "unset" label, and furthermore,
            \item All item values corresponding to the specification item names are non-empty values (consider this condition true if no specification item names exist), and furthermore,
            \item The condition is satisfied (consider this condition true if the Condition is empty).
        \end{itemize}
        \item[Condition 2.] Modal is set to true. (The actual necessity of this command will be determined later.)
    \end{description}
\end{enumerate}

\item To generate new configuration commands, process Commands in the ToBe model's Config section following the same procedure as for configuration removal commands. However, in this procedure, replace all occurrences of "unset" in the previous removal command procedure with "set."

\item Process the footer command similarly to the header command.

\item Process the mode-before and mode-after commands as follows (Fig. \ref{fig_mode_process}).

\begin{enumerate}
    \item For device configuration commands with Modal set to true (Modal commands), those that have no “set” or “unset” labels on the item values associated with their corresponding Spec. Items and have no subcommands in the tree are considered redundant Modal commands and are removed from the tree.
    \item Perform a depth-first preorder traversal of the tree, adding mode-before command(s) immediately preceding the Modal command and mode-after command(s) immediately following it. When multiple mode-before or mode-after commands exist, maintain the sequence as defined in the device configuration command template.
\end{enumerate}
\item Generate the final device configuration procedure by re-creating the list in the order of nodes reached through depth-first preorder traversal of the tree, as shown in Fig. \ref{fig_proc_gen}.

\end{enumerate}

\section{Details of the Generated Device Configuration Procedures}
\begin{figure}[!tb]
        \centering
        \includegraphics[width=\hsize]{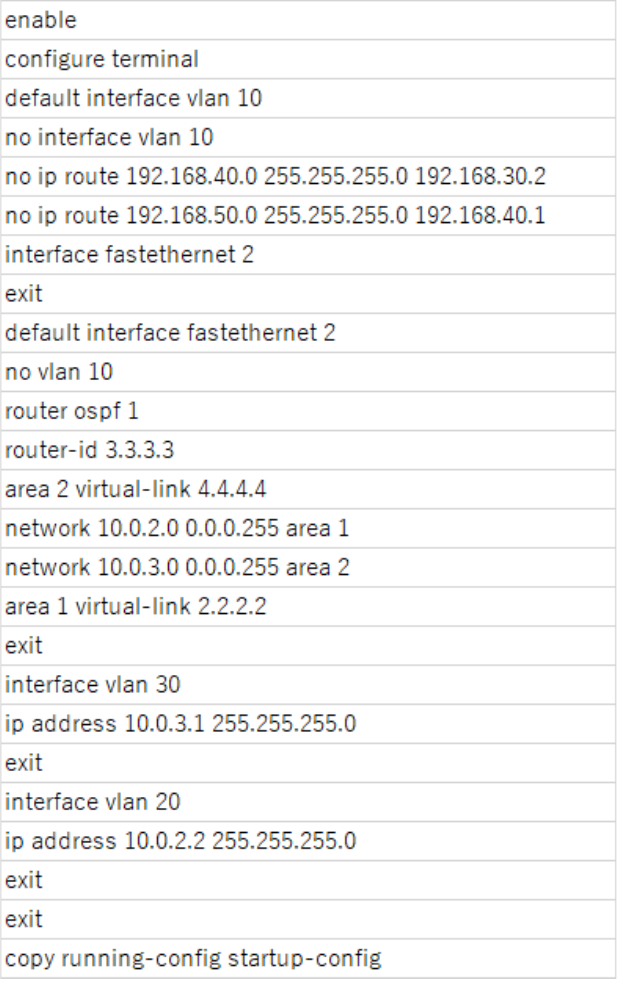}
        \caption{Generated configuration commands (for Cf1; partially)}
        \label{cf1_command}
\end{figure}

This section provides a detailed explanation of the generated device configuration procedures as described in Section \ref{sec_generate}. For brevity, only selected examples are presented. Focusing on the application case Cf1, when generating the device configuration procedure as shown in Fig. \ref{cf1_command}, it becomes clear that the generated output contains only the necessary configuration differences. Additionally, the presence of commands for configuration removal confirms that the system successfully detects and identifies obsolete information from the differences. The generated device configuration procedures can be executed from each network device's console mode (by copying and pasting) to apply the required settings to the ToBe network.

\begin{figure*}[t]
\centering
\begin{lstlisting}[basicstyle=\ttfamily\small,captionpos=b,frame=single]
C    192.168.30.0/24 is directly connected, Vlan30
C    192.168.10.0/24 is directly connected, Vlan10
S    192.168.40.0/24 [1/0] via 192.168.30.2
C    192.168.20.0/24 is directly connected, Vlan20
S    192.168.50.0/24 [1/0] via 192.168.40.1
\end{lstlisting}
\jecaption{Routing table of campus1 in the AsIs network}{list1}
\end{figure*}

\begin{figure*}
\centering
\begin{lstlisting}[basicstyle=\ttfamily\small,captionpos=b,frame=single]
O IA    10.0.8.0 [110/3] via 10.0.2.1, 01:52:25, Vlan20
O IA    10.0.9.0 [110/2] via 10.0.2.1, 01:52:25, Vlan20
C       10.0.2.0 is directly connected, Vlan20
C       10.0.3.0 is directly connected, Vlan30
O       10.0.1.0 [110/2] via 10.0.2.1, 01:52:25, Vlan20
O IA    10.0.6.0 [110/4] via 10.0.3.2, 01:52:45, Vlan30
O IA    10.0.7.0 [110/4] via 10.0.2.1, 01:52:25, Vlan20
O IA    10.0.4.0 [110/2] via 10.0.3.2, 01:52:45, Vlan30
O IA    10.0.5.0 [110/3] via 10.0.3.2, 01:52:46, Vlan30
\end{lstlisting}
\jecaption{Routing table of campus1 in the AsIs network}{list2}
\end{figure*}

\begin{figure*}
\centering
\begin{lstlisting}[basicstyle=\ttfamily\small,captionpos=b,frame=single]
campus4#traceroute 10.0.1.2
...
  1 10.0.6.1 0 msec
  2 10.0.7.1 0 msec
  3 10.0.8.1 0 msec
  4 10.0.9.1 0 msec
  5 10.0.1.2 0 msec 0 msec *
\end{lstlisting}
\jecaption{Routing information from campus4 to top in the ToBe network}{list3}
\end{figure*}

\begin{figure*}
\centering
\begin{lstlisting}[basicstyle=\ttfamily\small,captionpos=b,frame=single]
campus4#traceroute 10.0.1.2
...
  1 10.0.5.1 0 msec 0 msec 0 msec
  2 10.0.4.1 0 msec 0 msec 0 msec
  3 10.0.3.1 0 msec 0 msec 0 msec
  4 10.0.2.1 0 msec 0 msec 0 msec
  5 10.0.1.2 4 msec 0 msec *
\end{lstlisting}
\jecaption{Routing information from campus4 to top in the ToBe network when the link failure}{list4}
\end{figure*}

\section{Evaluation}
Using the networks shown in Fig. \ref{fig_asis_nw} and Fig. \ref{fig_tobe_nw} as application cases, we evaluated the effectiveness of the proposed method by assessing its ability to generate the expected device configuration procedures. For this evaluation, we developed a prototype tool implementing the proposed method in Java, referencing the algorithms detailed in Section \ref{sec_sabun} and \ref{sec_generate}. Additionally, this method utilizes astah \cite{astah} for creating models, which provides APIs for extracting model information from Java programs. By leveraging these APIs, we obtained the necessary information for detecting model differences and applied labels according to the algorithmic approach. Using this information along with the device configuration command templates, we successfully generated executable configuration procedures. Furthermore, to verify the correctness of the generated device configuration procedures, we conducted actual hardware-based validation. In this process, we received assistance from one network operations management expert, who verified two key aspects: the accuracy of routing tables and the proper implementation of route changes during network disconnections.

\subsection{Application Case Network Configuration}
We detail the configuration of the application case network used in this evaluation. The specific application case presented here represents a simplified model of inter-campus interconnections.

\subsubsection{Pre-modification AsIs Network}
The As-Is network before the changes (Fig. \ref{fig_asis_nw}) reproduces the network that was previously operated at Shinshu University. Static routing is used for routing, with routes configured manually, and the operational setup does not consider redundancy. Therefore, if a network failure occurs at a single point, communication from the affected site will be completely disrupted. The names of the network devices (name in  Hostname) are assigned from campus1 to  campus6. The routing table information obtained from  campus1 is shown in List \ref{list1}. Here,  C denotes  connected, indicating a directly connected route, while  S denotes  static, representing a manually configured static route.

\subsubsection{Post-modification ToBe Network}
The post-modification ToBe network (Fig. \ref{fig_asis_nw}) replicates the current network configuration in operation at Shinshu University. It utilizes a combination of dynamic and static routing, primarily employing the OSPF protocol for dynamic routing, resulting in a redundant configuration with multiple areas. This design ensures that when a network failure occurs at any single point, the network automatically reroutes traffic through alternative paths. The names (hostnames) of each network device are designated as campus1 through campus7, dc (data center), and top (entry/exit points).

\subsection{Evaluation of Device Configuration Procedures}
We detail the evaluation of the automatically generated device configuration procedures. While multiple valid command sequences could be assumed for a single configuration change, this evaluation assesses whether: (1) the procedures are acceptable to network operations experts, and (2) they function as expected when applied to the actual network infrastructure (AsIs configuration) through routing table analysis and path verification. Regarding (2) specifically, we performed path verification during network disconnections using the ICMP protocol \cite{tcpip} to confirm proper operation. The expected operational behavior is defined as meeting two conditions: complete absence of packet loss between each pair of network devices (as shown in Fig. \ref{fig_tobe_nw}), and successful retrieval of routing information (path information between devices in Fig. \ref{fig_tobe_nw}).

\subsection{Preparation}
The specific input data and devices used in this evaluation are as follows:
\begin{enumerate}
    \item AsIs model (Fig. \ref{fig_asis_model})
    \item ToBe model (Fig. \ref{fig_tobe_model})
    \item Device configuration command template (Fig. \ref{fig_template})
    \item Cisco 1812-J routers × 7 (Version 12.4(15)T11)
    \item Cisco 892 routers × 2 (Version 15.1(4)M5)
\end{enumerate}

The AsIs and ToBe models were created using the astah modeling tool, while the device configuration command template was expressed in CSV format.

The following outlines the experimental procedure:

\begin{enumerate}
    \item The experimenter inputs the input data into the proposed tool to generate device configuration procedures
    \item A network operations expert verifies whether the generated configuration procedures contain any issues
    \item The experimenter applies the generated configuration procedures to actual hardware to build the ToBe network
    \item Under the supervision of a network operations expert, the experimenter verifies expected behavior by examining routing tables and path verification during network disconnections
\end{enumerate}

\subsection{Results}
After issuing the generated device configuration procedures to actual hardware and verifying routing tables along with path verification using traceroute commands, while also intentionally causing network disconnections to confirm whether the correct paths to target network devices were displayed, we confirmed normal network operation. However, during configuration procedure generation, when searching for specification item groups with Config as the root node, no exploration order constraints were imposed between specification item group values that serve as Config's child nodes. As a result, some executions produced rearranged command sequences (e.g., vlan and hostname), though we confirmed that this did not affect functionality in this specific case.

\subsubsection{ToBe Network Routing Tables}
We issued the actual device configuration procedures output for each network device in the AsIs network and for the newly added network devices. The routing table obtained from campus1 during this operation is shown in List \ref{list2}. Here, "O" represents path information obtained through LSA Type1 and Type2, while "O IA" represents path information obtained through LSA Type3. Since both configurations apply the OSPF protocol, it can be seen that path information is successfully obtained even between different areas.

\subsubsection{Routing Table Behavior Following Network Disconnection}
    To verify that the OSPF protocol was functioning correctly, we intentionally caused network disconnections and checked the path information. The traceroute command issued from campus4 toward top is shown in List \ref{list3}. Here, for all paths, we confirmed that they fully cover the route from campus4 to top, verifying that there are no operational issues. Next, in List \ref{list4} we show the path behavior when disconnecting the network between (campus4) and campus5 (10.0.6.0/24). This demonstrates that the system correctly determines an alternate route, bypassing the disconnected network (10.0.6.0/24), confirming the expected behavior.

\subsection{Discussion}
Based on the evaluation results, the successful generation of network device configuration procedures that matched expectations demonstrates the method's demonstrated effectiveness. This demonstrates that the proposed approach effectively identifies model-specific differences and, through device configuration command templates, properly derives the necessary configuration commands.

Furthermore, the automated generation of device configuration procedures offers several benefits: preventing engineer oversights and misinterpretations, identifying discrepancies in configuration documentation standards among engineers, and reducing the burden on engineers by generating configuration procedures during the design phase. However, we do not address the additional burden placed on engineers required to create these models. Notably, duplicate specification values may lead to duplicate IP addresses in the same network or to models containing invalid values, resulting in the generation of inappropriate configuration procedures. To mitigate these risks, we propose developing:

- Static verification \cite{Onishi01} tools to enable model validation during the creation process, making it easier for engineers to create models

- Dynamic verification \cite{satake} tools using network simulators to validate models after creation

While our evaluation has focused on features such as ACLs, OSPF protocol, and VLANs, many other functionalities, protocols, and proprietary protocols from various vendors exist. Therefore, future work should involve further expanding the proposed method and evaluating its effectiveness. Additionally, we have not evaluated whether the method can flexibly adapt to imposing constraints on the exploration order of child nodes in Config configurations or generate configuration procedures according to engineers' specific policies.

Regarding ACL configuration changes, we note that these depend on the model created during initial configuration. When modifying ACL settings using this method, do not alter existing ACLs. Instead, describe ACL-related information within the model and generate device configuration procedures. The ACL is then updated by applying the newly created configuration. However, when modifying these ACLs, the appropriate approach depends on the specific circumstances of the change request and may vary based on individual engineer policies. Therefore, implementing a configuration procedure generation system that can modify existing ACLs and allow selection based on engineers' policies would be preferable.

\section{Conclusion}
This paper proposes an automated method for generating device configuration procedures based on network configuration models. The proposed method generates device configuration procedures (sequences of configuration commands) by analyzing differences between two network configuration models. Through evaluation using actual operational networks, we confirmed that the proposed method successfully generates expected configuration procedures, verifying its effectiveness.

Future research directions include expanding the proposed method to enable flexible generation of configuration procedures according to engineers' policies, while also conducting both quantitative and qualitative evaluations of its effectiveness. Additionally, we need to assess the proposed method's versatility for generating device configuration commands in multi-vendor environments and evaluate its effectiveness in real-world networks using various unaddressed protocols. Furthermore, we plan to expand the method to support automated generation of device configuration procedures and explore integration methods with conventional approaches. First, we will focus on enhancing the method's expressive capabilities, investigating the method's current limitations in application scope, and conducting surveys among various engineers regarding configuration procedure creation policies.

The time required for creating the proposed model and its validity evaluation also remain future challenges. We also intend to consider integration with network auto-generation methods that combine Ansible and NETCONF, which are recognized as effective network automation approaches.

\section*{Acknowledgments}
We express our gratitude to NTT EAST for their cooperation in managing and operating Shinshu University's network, which enabled the conduct of this research. This study was supported by JSPS KAKENHI Grant Number JP17KT0043.

\bibliographystyle{IEEEtran}
\bibliography{tecrep}

\end{document}